# Unsupervised Denoising of Optical Coherence Tomography Images with Dual-Merged Cycle-WGAN

Jie Du Xuanzheng Qi Xujian Yang Kecheng Jin Hu Chen

*Abstract*—**Nosie is an important cause of low quality Optical coherence tomography (OCT) image. The neural network model based on Convolutional neural networks(CNNs) has demonstrated its excellent performance in image denoising. However, OCT image denoising still faces great challenges because many previous neural network algorithms required a large number of labeled data, which might cost much time or is expensive. Besides, these CNN-based algorithms need numerous parameters and good tuning techniques, which is hardware resources consuming. To solved above problems, We proposed a new Cycle-Consistent Generative Adversarial Nets called DualMerged Cycle-WGAN for retinal OCT image denoiseing, which has remarkable performance with less unlabeled traning data. Our model consists of two Cycle-GAN networks with imporved generator, descriminator and wasserstein loss to achieve good training stability and better performance. Using image merge technique between two Cycle-GAN networks, our model could obtain more detailed information and hence better training effect. The effectiveness and generality of our proposed network has been proved via ablation experiments and comparative experiments. Compared with other state-of-the-art methods, our unsupervised method obtains best subjective visual effect and higher evaluation objective indicators.**

## 1. Introduction

OCT is a technique proposed by Huang *et al* [1], which is for high-resolution tomography of the internal microstructure of biological tissue based on the low coherent properties of light. The technique uses the Michaelson interferometer to complete coherent selection, spatially 2D or 3D scanning of biological tissue, which is a high-speed tomography technique for non-intrusive biological tissue [2]. At present, not only in ophthalmology, dentistry and other clinical diagnosis, but also in the field of industrial testing, this technique has been widely applied. However, OCT is a high- resolution imaging technology, which requires stable signal acquisition process. It will inevitably receive noise pollution, while causing structure blur and distortion, adversing to the accurate judgment of subsequent images. Specifically, OCT imaging is based on coherent detection technology, which makes noise the primary cause of noise [3]. Although image technology and related equipment has been continuously developing, the problem of noise has not been solved well and it has seriously affected the automatic diagnosis of OCT images, such as registration [4], retinal lesion region segmentation [5], and retinal layer information analysis [6]. Therefore, how to denoise OCT image is a primary task to improve the performance of automatic diagnostic performance. Recently, many methods have been proposed for OCT image denoising. These methods can be divided into two aspects: hardware based and software based. Hardware based algorithms is mainly about improvements to the imaging system. However, these algorithms is not quite useful because it require specially designed acquisition systems and thus not suitable for commercial useage. Software based algorithms is mainly about processing of digital image signal after OCT imaging. In the traditional software based methods, wavelet transform-based methods [7] are widely used for OCT image denoising. These methods decompose OCT image into images in different frequency bands by wavelet transform, in which the colored noise is distributed in the high frequency component and the white noise is distributed in the low frequency component, The high frequency details are omitted or weighted, and the clear image can be synthesized after the image components are reconstructed. After that, A new wavelet transform method based on the combination of wavelet transform and Wiener filter [8] is proposed. This method decomposes the image into four different frequency bands by wavelet transform, and does not change in the low-frequency part, and uses Wiener filter in the high-frequency part. However, these methods show a certain degree of overfitting. Bo and Zhu [9]proposed wavelet modification based block matching and 3D filtering (BM3D) for OCT image denoising. After combining the advantages of traditional spatial domain and transform domain denoising algorithms, a DDID algorithm is proposed [10], which can effectively remove the additive Gaussian white noise of graphics. There are some other methods to deal with this noise. The total variation approximation method is applied to the multiplicative noise model [11]. This method uses a constrained optimization approximation with two Lagrange multipliers to build the model, but the fitting term is non convex, So Yang et al. Used the first- order primal dual algorithm [12] to deal with the image with speckle noise, the result is very good to retain the image details, and the effect is better than the total variation method. However, these methods have some shortcomings such as can't capture enough image features, difficult to choose the right thresholds.

With the development of deep learning, the method based on convolutional neural network [13] has greatly improved the image denoising, in which the stacked sparse autoencoder is applied to natural image denoising [14]. Hu Chen *et al* Proposed a residual encoding decoding for low- dose CT image denoising CNN [15]. Ma *et al* considered image denoising as the problem of image to image conversion, and proposed Speckle noise reduction in optical coherence tomography images based on edge-sensitive CGAN [16] . However, All of the above deep learning methods belong to supervised learning, and all of them need labels corresponding to images to carry out experiments.

To solve above problems, we proposed a newly unsupervised method which based on our proposed double-model Cycle-Consistent Adversarial Networks together with image merge(fusion) called Dual-Merged Cycle-WGAN. This method can learn a lot of image mapping from SD-OCT to EDI-OCT(enhanced depth imaging optical coherence tomography) unsupervisedly and can achieve remarkable performance on denoising OCT image only using a small mount of unlabeled image data.

## 2. Related Work

### 2.1. CycleGAN

CycleGAN is an image-to-image translation based on GAN, which defined generator: $G: X \to Y$ and $F: Y \to X$ discriminator $D_X$ distinguishes between $x$ and $F(y)$, and $D_Y$ distinguishes between $y$ and $G(x)$. CycleGAN introduces the idea that "if we translate from one domain to another and back again we should arrive where we started" [17] . The objective function of CycleGAN consists of two types of loss. The adversarial loss evaluates the distance between the distribution of the generated image and the real image. Cycle consistency loss defined as $F(G(X)) \approx X$, and $G(F(X)) \approx y$, which calls the cycle consistency. Based on the cycle consistency, only the source data set and the target data set need to be used in the training process, and there is no need to have a one-to-one mapping relationship between the data, which can solve the problem of not being able to obtain or difficult to obtain a paired data set.

In CycleGAN, we have two adversarial losses:

$$L_{GAN}(G, D_Y, X, Y) = E_{y \sim pY(y)}[\log D_Y(y)] + E_{x \sim pX(x)}[\log(1 - D_Y(G(x)))]$$

and

$$L_{GAN}(F, D_X, Y, X) = E_{x \sim pX(x)}[\log D_X(x)] + E_{y \sim pY(y)}[\log(1 - D_X(F(y)))]$$

The cycle consistency loss is defined by

$$L_{cyc}(G,F) = E_{x \sim pX(x)}[\|F(G(x)) - x\|_1] + E_{y \sim pY(y)}[\|G(F(y)) - y\|_1]$$

Combining the adversarial losses and the cycle consistency loss, we obtain the full objective function:

$$L(G, F, D_X, D_Y) = L_{GAN}(G, D_Y, X, Y) + L_{GAN}(F, D_X, Y, X) + \lambda L_{cyc}(G, F)$$

where $\lambda$ controls the relative importance of the cycle consistency loss. In the training phase, the parameters in $G, F, D_X$, and $D_Y$ are estimated by optimizing the full objective function and we get

$$G^*, F^* = \arg \min_{G,F} \max_{D_X, D_Y} L(G, F, D_X, D_Y)$$

CycleGAN as an image conversion method has important applications in fields such as photo enhancement, image coloring, style transfer, etc.

### 2.2. WGAN

GAN has achieved great success in image translate after it been put forward. However, it has problems such as difficulty in training and insufficient diversity of generated results. Wasserstein GAN(WGAN [18]) made some improvement of the problems of GAN.

Arjovsky *et al*. stated that the difficulty in trainnning of GAN is due to the poorly designed loss function. Many loss functions commonly used in GAN, such as JS divergence, are locally saturated, which will cause the problem of gradient disappearance. Therefore, they proposed the Wasserstein distance with better continuity and differentiability.

Suppose the distribution of the real image and the generated image are $P_r$ and $P_g$. The Wasserstein distance between $P_r$ and $P_g$ defined by

$$W(P_r, P_g) = \sup_{\|f\|_L \le 1} E_{x \sim P_r}[f(x)] - E_{x \sim P_g}[f(x)]$$

supremum is taken over all 1 — Lipschitz function $f: X \to R$ and x is a tight metric space. In the context of GAN, function $f$ Corresponds to discriminator D(x) and objective function of WGAN becomes:

$$\min \max E_{x \sim P_r}[D(x)] - E_{x \sim P_g}[D(X)]$$

where D is the set of 1 — Lipschitz functions and $P_g$ is the model distribution defined by x = G(z), z ~ p(z), where p(z) is some simple noise distributions, such as uniform distribution or Gaussian distribution. The 1-Lipschitz constraint on the discriminator is through clipping. The weight of the discriminator is located in a compact space [-c, c].

In the final algorithm, WGAN has four changes relative to GAN:

- Remove sigmoid in the last layer of the discriminator

- The loss of generator and discriminator does not take log
- Every time the parameters of the discriminator are updated, their absolute value is truncated to no more than a fixed constant c
- Replaces the optimizer Adam with RMSProp

# 3. Methods

## 3.1. Overcview

In this article, we proposed a newly unsupervised method which based on our proposed Dual-Merged Cycle- WGAN for OCT image denoiseing, which has remarkable performance with less unlabeled traning data. We use two Cycle-GAN networks combining with image fusion technique. Specifically, let us assume the set of original noise and clean OCT images are $X_*$, $Y_*$, two Cycle-GAN networks $M_i$, $M_2$, and their generator is $G_i$, $G_2$ respectively. Fisrt, we randomly crop original noise and clean images into small pieces, each consisting of 2100 pictures. Then, apply normalization on these data. Denote these processed data by $X$ for noise and $Y$ for clean. We use $X$ and $Y$ to train our first model $M_i$ . The output of the $M_i$ is merged with original image via linear combination and plus samesized random img which generate from standard normal distribution. The result above is taken as input of $M_2$ and the output of $M_2$ is our predict image. More formally, let $x$ *e* $X$, the computation procedure is described as followed:

$$x \longrightarrow G_i(x) \longrightarrow aG_i(x) + bx = x_i$$

$$x_i \longrightarrow x_i + z = x$$

$$x_2 \gg G_2(x_2)$$

Where $a = 0.8$, $b = 0.2$, $z$ — $N(0,1)$, $y = 6_2(x_2)$ is the output of $M_2$. We use $x_2$ and y to train our model $M_2$, and take y as final predict result.

We also imporve the stucture of two Cycle-GAN networks.

- Improve the original U-Net with newly Multi-U-Net
- Use the Wasserstein loss instead of the orignal loss of Cycle-GAN network

To evaluate denoising performance, we use four evaluation indicators: SNR, ENL, PSNR, SSIM.

## 3.2. Data Augmentation

Our experiments use the dataset which contains 21 noised pictures and corresponding clean pictures(each of size 360 x 800 and clean pictures only for metrics) which is not enough for the training of Our Network. Hence, we preprocessed and augmentation our dataset by the following steps:

**3.2.1. Adjustment for the contrast.** By Adjusting the contrast of the clean pictures, we can sharpen the edge of lines , which is more conducive to clarity the original pictures.

**3.2.2. Magnify.** To get more detailed features, improve network efficiency and generalize model's performance, we magnify our orginal image by three times.

**3.2.3. Cut.** The original images is too large but the amount is too small, which is not suitable for network's training. Therefore, we uniformly cut the image into 256 x 256 images, which generate 2100 images for our model training.

## 3.3. Dual-Merged Cycle-WGAN

Our model consists of two Cycle-GAN networks with imporved generator and descriminator which both using Multi-U-Net.

## 3.4. Multi-U-Net: a generator for improving Cycle-GAN

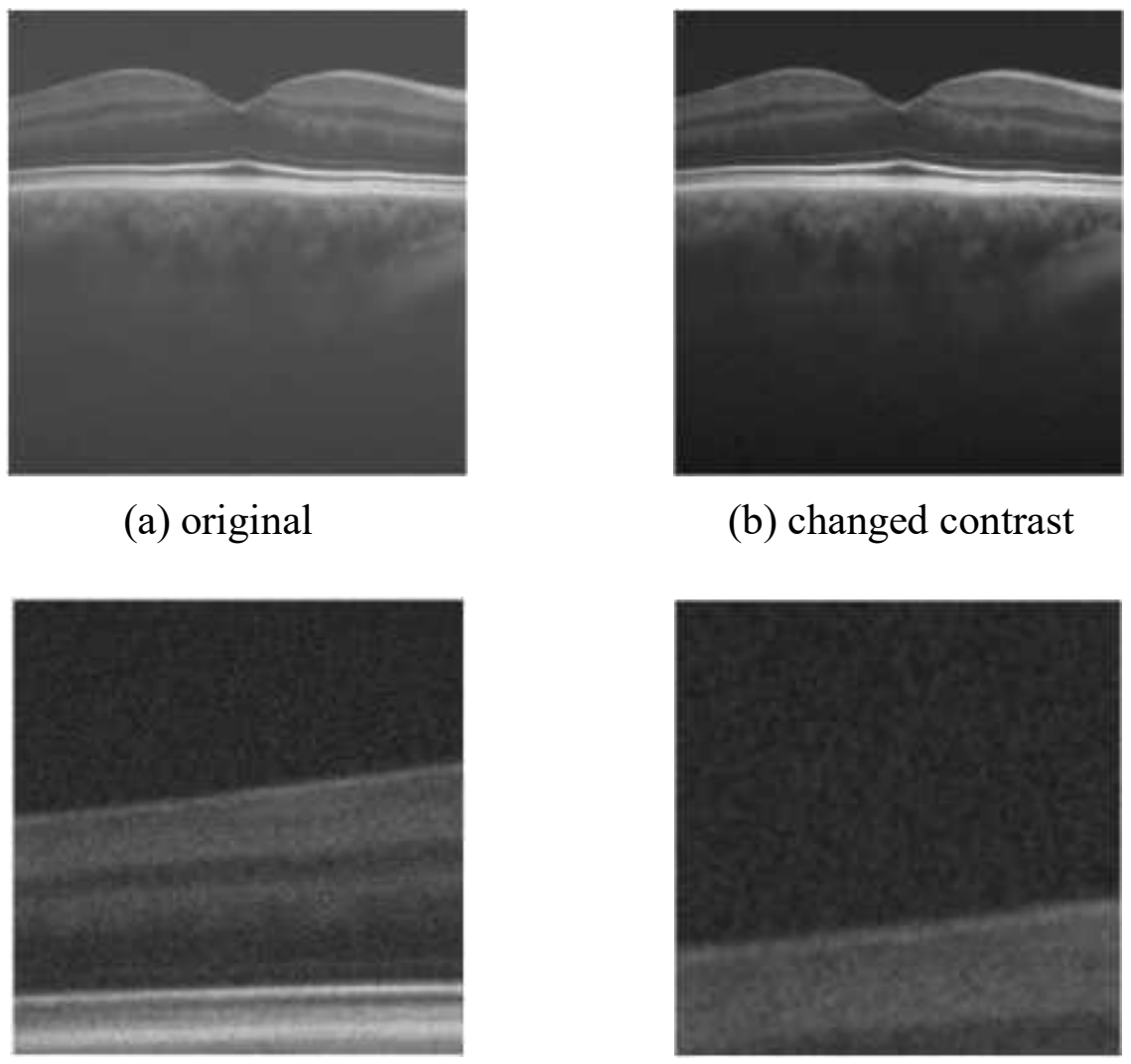

(a) original (b) changed contrast

(c) original (d) magnify

U-Net [19] is a structure commonly used in the field of medical image processing, which is also used by the generator in the normal Cycle-GAN. The structure of U- Net make it easy to capture the detailed information behind images. Moreover, U-Net can localize the segmented parts of the image. Motivated by these good properties of U-Net, we proposed one with a bit more complex structure called Multi-U-Net. First, divide the input into multiple parts, put them into multiple U-net with different depth in turn. The input image is divided with different scalar according to different depth of downsampling. After that, data are trained through the block layer. And then we use upsampling to unify the scales of different scales of the block data to the same size as the input image. In this way, different depths of this U-Net Structure can learn image features with different dimensions. Finally merging the above results of multiple U-Nets to get the Multi-U-Net results. The visual procedure can be found in figure 1 The results are combined with the hideen feature information of multiple dimensions, and thus better noise reduction performance. Nesides, increasing the number of U-Net with different depths is similar to increasing the width of a single layer of one fully connected network. In this way, we can enhance the network's image learning ability.

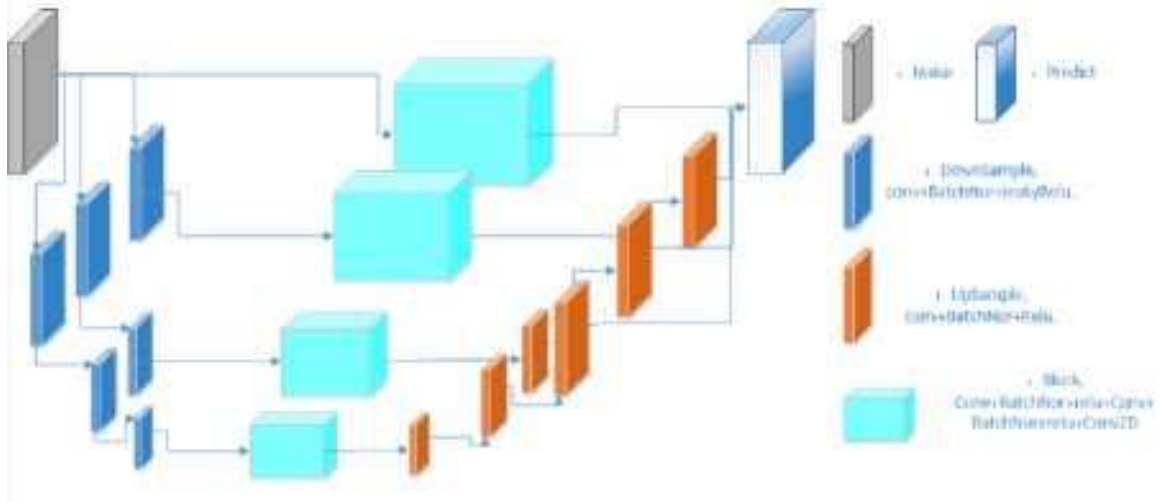

Figure 1: MultiU-Net Structure

For better denoising performance, we introduce the image merge technique between two Cycle-GAN networks.

## 3.5. Multilevel Cycle-GAN: structure of network for training

As previously discussed, single Cycle-GAN can achieve a better noise reduction effect. However, Cycle-GAN is a Image-to-Image-Translation network [17], that is, we can use Cycle-GAN to convert noise images into clear images, or input clear images into the network to get noise images, The greater the difference between the noise image and the clear image, the more difficult the network training will be, and the worse the effect will be. Thus, our model use image merge between two Cycle-GAN networks to obtain more detailed information and hence better training effect.

The figure 2 below is part of the structure we designed.

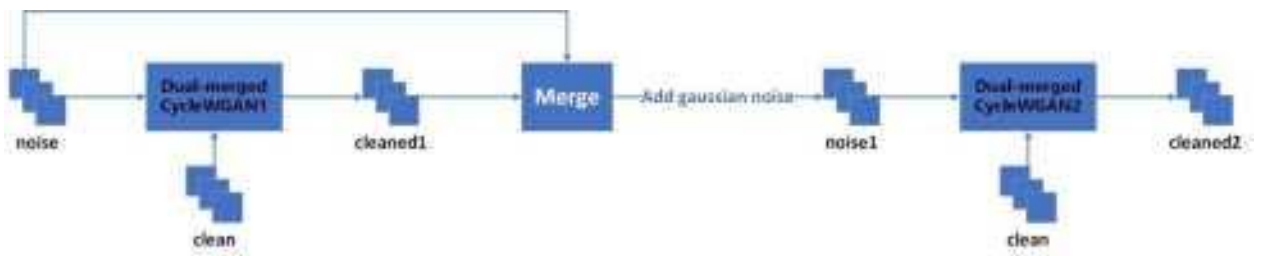

Figure 2: Iamge merge. (a)noise: original img, (b)clean: target img, (c)clean1: ouput of $G_1$, (d)noise1: after merging, (e)clean2: ouput of $G_2$

## 3.6. Loss function

To improve the training stability of Cycle-GAN, we extended the idea of WGAN and improved WGAN to Cycle- GAN.

**3.6.1. Cycle-GAN.** In Cycle-GAN, we have two adversarial losses:

$$L_{GAN}(G, D_Y, X, Y) = E_{y\sim p_Y(y)}[\log D_Y(y)] + E_{x\sim p_X(x)}[\log(1 - D_Y(G(x)))]$$

and

$$L_{GAN}(F, D_X, Y, X) = E_{x\sim p_X(x)}[\log D_X(x)] + E_{y\sim p_Y(y)}[\log(1 - D_X(F(y)))]$$

The cycle consistency loss is defined by

$$L_{cyc}(G, F) = E_{x\sim p_X(x)}[\|F(G(x)) - x\|_1] + E_{y\sim p_Y(y)}[\|G(F(x)) - y\|_1]$$

Combining the adversarial losses and the cycle consistency loss, we obtain the full objective function:

$$L(G, F, D_X, D_Y) = L_{GAN}(G, D_Y, X, Y) + L_{GAN}(F, D_X, Y, X) + \lambda L_{cyc}(G, F)$$

where $\lambda$ controls the relative importance of the cycle consistency loss. In the training phase, the parameters in $G, F, D_X$, and $D_Y$ are estimated by optimizing the full objective function and we get

$$G^*, F^* = \arg\min_{G,F}\max_{D_X, D_Y} L(G, F, D_X, D_Y)$$

**3.6.2. Cycle-WGAN.** The proposed adversarial losses for Cycle-WGAN are

$$L_{WGAN}(G, D_Y, X, Y) = E_{y\sim p_Y(y)}[D_Y(y)] - E_{x\sim p_X(x)}[D_Y(G(x))]$$

and

$$L_{WGAN}(F, D_X, Y, X) = E_{x\sim p_X(x)}[D_X(x)] - E_{y\sim p_Y(y)}[D_X(F(x))]$$

Combined with cycle consistency loss, our full objective for CycleWGAN is:

$$L(G, F, D_X, D_Y) = L_{WGAN}(G, D_Y, X, Y) + L_{WGAN}(F, D_X, Y, X) + \lambda_0 L_{cyc}(G, F)$$

# 4. Experiments

## 4.1. Experimental data

The study is approved by the Institutional Review Board of Sichuan University, and informed consent was obtained from all subjects. As mentioned before, our experiment is based on 21 noise and 21 clear images of SD-OCT with a size of 360 x 800 pixels. For the comparative experiment, we use the method in section 3.5 to divide 21 pictures into 2100 pictures with 256 x 256 pixels, of which 1890 are used as the training set and 210 as the test set. All subsequent comparative experiments use this data set to ensure the fairness of the experiment. For ablation experiments, we use the same data expansion method to generate 2500 training sets and 500 test sets. This data set is used in each group of ablation experiments. Data expansion is one of the effective strategies to increase the diversity of data distribution and alleviate the over fitting problem. In this paper, we expand the data set by enlarging the pixels of the picture and then randomly cutting. Increased the amount of data by 100 times.

## 4.2. Evaluation Metrics

In order to evaluate the denoising performance of different methods, four indexes including Signal-to-noise ratio (SNR), Equivalent numbers of looks (ENL), Structural similarity index measure (SSIM) and Peak signal-to-noise ratio (PSNR) were used to analyzed experimental results quantitatively. These four indicators are calculated as followed.

**4.2.1. SNR.** the Signal-to-noise ratio(SNR) is a global performance measure which has been widely used to evaluate denoising performance when the reference clean images are not available [20] [21].
The SNR can be calculated as:

$$SNR = 20 - \log[\max(I)^2/a^2]$$

where *max (I)* is the maximum possible pixel value the denoised image, *a* is the standard deviation of noise in the background region.

**4.2.2. ENL.** Equivalent numbers of looks(ENR) [22], is a commonly used performance measure for speckle suppression, which measures smoothness in regions that appear to be homogeneous which is defined as:

$$ENL = \frac{乌}{a^2}$$

where 甘 and *a* denote mean value and standard deviation of the background region, respectively.

**4.2.3. SSIM.** The structural similarity index measure (SSIM) is a method for predicting the similarity of two picture [23]
It's formula is based on three comparison measurements between the two picture:x and y:

$$l(x,y) = \frac{2 \ 翊 \ c_i}{产 X \ 2}$$

$$c(x,y) = \frac{a_x a_y \ c_1}{+ \ +}$$

$$s(x,y) = \frac{a_X \ a_y \ c_i}{a_x a_y + c_3}$$

SSIM is then a weighted combination of those comparative measures:

$$SSIM(x, y) = [l(x, y)^a - c(x, y)' . s(x, y)^y]$$

where *a, MY* > 0 are constant, 皿, 驯 and $a_x$, $a_y$ are the average and variance of x and y. $a_{xy}$ is the covariance of x and y, $c_1$, $c_2$, $C_3$ are constant too.
The larger the value of the SSIM, the higher the similarity between the two pictures.

**4.2.4. PSNR.** Peak signal-to-noise ratio (PSNR) is an engineering term for the ratio between the maximum possible power of a signal and the power of corrupting noise that affects the fidelity of its representation [24].
Given a noise-free I with size of m x *n* and its noisy approximation

$$\frac{1}{mn} \sum_{i=0}^{m-1} \sum_{j=0}^{n-1}$$

*K*, MSE is defined as:

$$MSE \qquad [I(i,j) - K(很)]^2$$

The PSNR is defined as:

$$PSNR = 10 .\%(\frac{MAX_I^2}{MSE})$$

where MAX/ is the maximum possible pixel value of the image (When the pixels are represented using 8 bits per sample, this is 255)

## 4.3. Qualitative Evaluation

As can be seen from figure 3, the proposed unsupervised learning method performs well in all test samples, eliminating image noise from different regions. Besides, the proposed model retains and enhances retinal layer structure and choroidal vessels. Although good results are obtained in undamaged images, our method can also have good results for some abnormal parts.

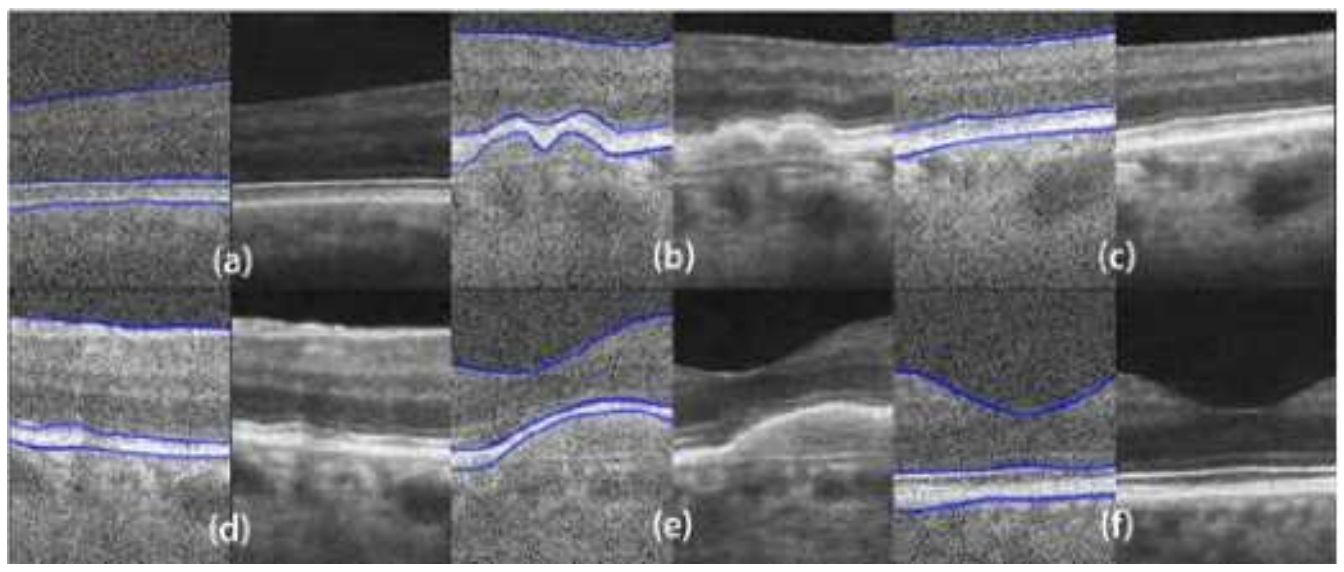

Figure 3: The denoising result of our method. left: original OCT, right: denoised OCT

In order to further evaluate the performance of our proposed method, figure 4 shows examples of denoising results of different methods, in which all the proposed methods use the same training set and test set. It can be seen that although BM3D method can remove noise well, the boundary of retinal layer is blurred. Moreover, Noise2Noise method remove noise well, it can be seen that the layers are not smooth enough and there are still many artifacts. Pix2pix method can be seen that it performs poorly in the test results and can not well suppress these noise, which will also lead to blurred retinal layer structure. In the results of WGAN and CGAN, the edge of retinal layer is distorted, while in the results of CGAN, the external limiting membrane (ELM) is not well enhanced. Compared with these methods, our proposed method removed most noise and enhanced retinal layer information and make its structure with clearer stratigraphic boundary, which proves the effectiveness of the proposed method.

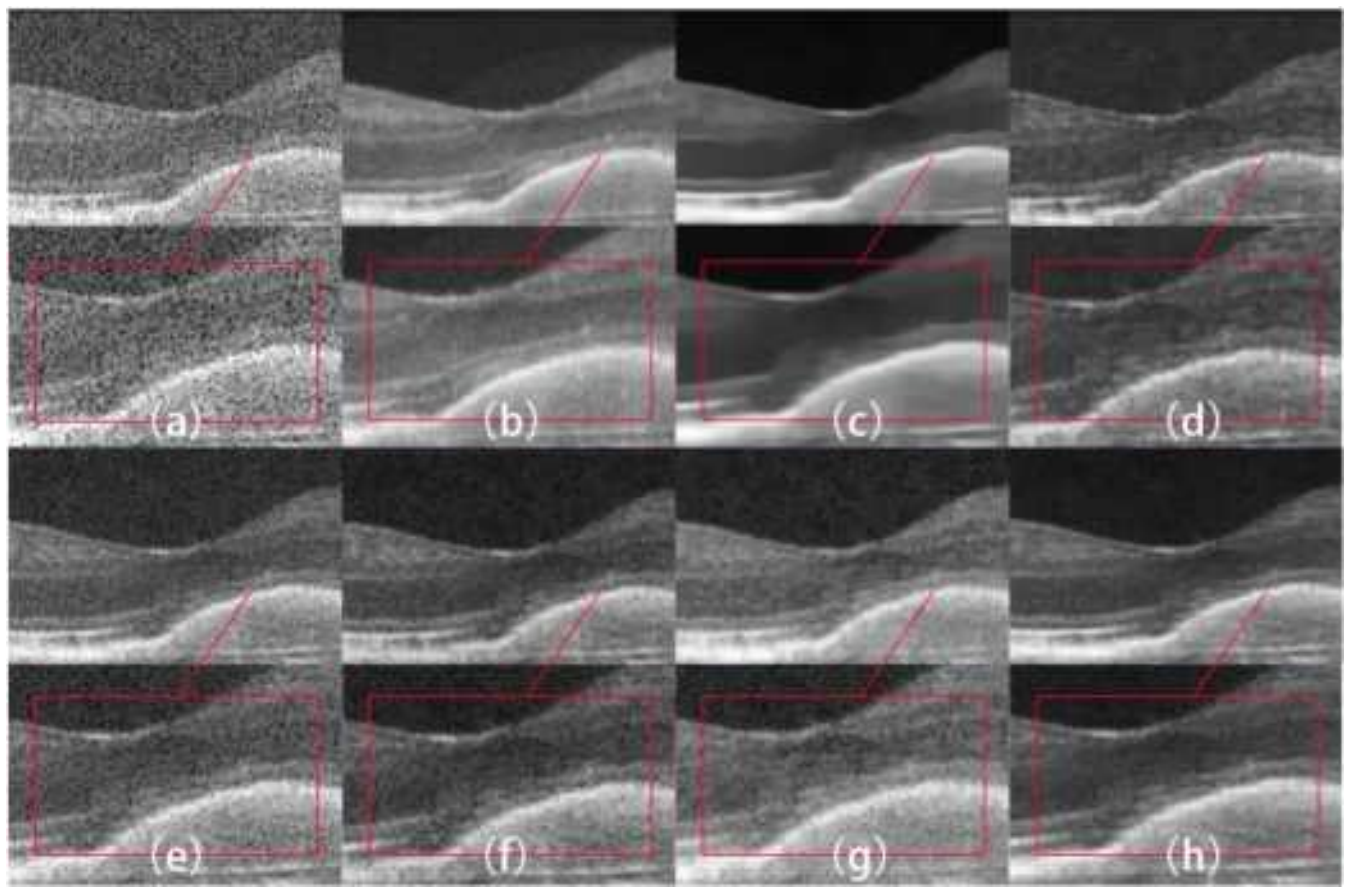

Figure 4: One example of denoising results with different methods. (a) Original image (b)Target image (c)BM3D (d)Noise2noise (e)Pix2Pix (f) WGAN (g)CGAN (h)Ours

| | SSIM | PSNR | SNR | ENL |
|---|---|---|---|---|
| BM3D | 0.9107 | 50.1429 | 59.0070 | 6.6789 |
| Noise2Noise | 0.9307 | 59.8852 | 27.1763 | 17.0659 |
| Pix2Pix | 0.9670 | 63.6405 | 20.5934 | 3.3614 |
| WGAN | 0.9718 | 66.1759 | 24.1671 | 2.2905 |
| CGAN | 0.9893 | 68.0705 | 23.4571 | 1.0895 |
| Dual-merged CycleGAN | 0.9994 | 69.7857 | 25.7563 | 1.3780 |

TABLE 1: Tabel C

## 4.4. Quantitative Evaluation

In order to quantitatively evaluate the speckle elimination performance, four indexes including SNR, ENL, SSIM and PSNR are listed in Table C.

The performance of some typical traditional methods is shown in the table. The structural similarity and peak signal-to-noise ratio of BM3D are low, and the SNR and ENL are much higher than those of the target image, which may be caused by its poor ability to suppress speckle noise. Noise2Noise has a good effect on SNR, but its performance on the other three indicators is very poor. This may be due to artifacts between retinal layers. The middle section lists the quantitative performance of some of the most advanced depth learning based methods, including Pix2Pix, WGAN and CGAN. Compared with other deep learning algorithms, CGAN obtains the lowest ENL, and also obtains good results in SSIM and PSNR. But his SNR is still far from the clear picture. Among them, pix2pix algorithm performs poorly in four indicators. Although WGAN works well on SNR and ENL and is very close to the target, its SSIM and PSNR are not the best. Compared with other methods, the Dual-merged-CycleWGAN proposed by us has achieved the best results in all four indicators.

## 4.5. Ablation Experiment

In order to evaluate the contribution of the loss function used in the proposed Dual-merged-CycleWGAN, four

| | SSIM | PSNR | SNR | ENL |
|---|---|---|---|---|
| UNet | 0.8992 | 66.6716 | 16.7546 | 1.8533 |
| Wassersein | 0.9389 | 67.1549 | 20.2985 | 1.9382 |
| Multi-UNet | 0.9796 | 69.3641 | 20.5140 | 1.9662 |
| double layer Cycle-GAN | 0.9293 | 68.3171 | 20.5809 | 1.9872 |
| Dual-merged Cycle-WGAN~~0.9995~~69.9221~~20.3607~~2.0168 | | | | |

TABLE 2: Tabel D

ablation experiments on the loss of original Cycle-GAN, the loss of original + Wasserstrein, the loss of MultiUNet and double-layer Cycle-GAN, and the loss of the proposed Dual-merged-CycleWGAN were carried out using the same training strategy. The following figure shows the original OCT image and the corresponding denoising results with different loss functions, and Table D shows the corresponding quantitative evaluation results:

It can be seen from Table D that the network with Wasserstrein loss function is compared with the original Cycle-GAN network. The performance of SSIM, PSNR, SNR and ENL has been improved. This may be that the Wasserstein distance has good smoothing properties compared with the JS distance of the original network, which can effectively solve the problem of gradient disappearance, make the denoising effect of the network better, the definition of the denoised picture higher and the structure retention more complete. Compared with the original Cycle- GAN network, the network using Multi UNet structure is almost consistent with the original results in SNR, but it is improved to a certain extent in PSNR, SSIM and ENL. This may be because the use of Multi UNet structure can retain more training parameters, make the mapping learned by the network more complex, and better deal with the denoising problem. A double layer Cycle-GAN network is used, which is compared with the original Cycle-GAN network. The four evaluation indexes have also been improved to a certain extent, which may be because the use of double layer Cycle- GAN process makes the relationship of network mapping simple, reduces the complexity of the network from noisy pictures to clear pictures, and improves the denoising effect. Finally, the proposed method is compared with the original Cycle-GAN network. Our proposed method is also superior to the original method in all indicators. The corresponding SSIM, PSNR, SNR and ENL are increased by 10.1%, 4.8%, 3.0% and 8.8% respectively. Compared with other previous methods, the best scores are obtained on all evaluation indexes except SNR. These results show the rationality of this method and the effectiveness of network structure design.

# 5. Conclusion and Future work

We proposed a new Cycle-Consistent Generative Adversarial Nets called Dual-Merged Cycle-WGAN for OCT image denoiseing, which has remarkable performance with less unlabeled traning data. This is the first time using Dual-Merged Cycle-WGAN for OCT image denoising and achieved a good denoising effect. Unlike previous neural network algorithms, we used a more complex-structure, which allowed our model to learn more hidden features of the image through a large number of hideen parameters without setting too many hyperparameters. In addition, our new proposed Dual-Merged Cycle-WGAN, compared with the previous algorithm, can get a good noise-cancelling effect only by training a small amount of noise images. Experiments results show that our network obtains good subjective visual effect and higher evaluation objective indicators, which make retinal layer edge

Still, there is some limitations in our study. Our data only have 21 noise pictures, which is not comprehensive enough. Although Our model performed well on our dataset , we believe if there is more datasets we can trained our model further and get better performance. Therefore, an important future research is to test our model on more actual OCT images and further optimize the hyperparameters of the proposed model to achieve better generalization capability.

Besides, how to speed-up the parallel computation of our model is also an important aspect that we will continuously focus on.

# Acknowledgments

The authors would like to thank the editors and all of the anonymous reviewers for their constructive feedback and criticisms.